\documentclass{PoS}

\usepackage{amsmath}							

\newcommand{\fig}[1]{Fig.~\ref{#1}}

\newcommand{\sect}[1]{Section~\ref{#1}}

\title{Exclusive $J/\psi$ and $\Upsilon$ photoproduction and the low x gluon}

\ShortTitle{Exclusive $J/\psi$ and $\Upsilon$ photoproduction}

\author{
\speaker{S.~P.~Jones}$^a$, A.~D.~Martin$^{b}$, M.~G.~Ryskin$^{bc}$ and T.~Teubner$^d$ \\
\llap{$^a$} Max Planck Institute for Physics, F\"ohringer Ring 6, 80805 Munich, Germany\\
\llap{$^b$} Institute for Particle Physics Phenomenology, Durham University, Durham DH1 3LE, U.K.\\
\llap{$^c$} Petersburg Nuclear Physics Institute, NRC Kurchatov Institute, Gatchina, St. Petersburg, 188300, Russia \\
\llap{$^d$} Department of Mathematical Sciences, University of Liverpool, Liverpool L69 3BX, U.K.
E-mail: \email{sjones@mpp.mpg.de}, \email{a.d.martin@durham.ac.uk}, \email{ryskin@thd.pnpi.spb.ru}, \email{thomas.teubner@liverpool.ac.uk}
}

\abstract{
We discuss the potential to constrain the small-$x$ PDFs using the exclusive production of heavy vector mesons. The calculation of $J/\psi$ and $\Upsilon$ photoproduction at NLO in collinear factorisation is described. The different behaviour of the NLO corrections for $J/\psi$ and $\Upsilon$ is highlighted and we outline what might be expected from the inclusion of these processes in a PDF fit.
}

\FullConference{XXIV International Workshop on Deep-Inelastic Scattering and Related Subjects\\
		 11-15 April, 2016\\
		 DESY Hamburg, Germany}

\begin{document}

\section{Introduction}

Currently there is a large uncertainty in the low $x \lesssim 10^{-3}$ behaviour of the gluon distribution in the global parton distribution function (PDF) analyses, particularly at low scales~\cite{Ball:2014uwa,Harland-Lang:2014zoa,Dulat:2015mca}. A wide variety of data on exclusive heavy vector meson (HVM) production $\gamma^* p \rightarrow V + p$ and ultraperipheral HVM production $pp \rightarrow p + V + p$, with $V=J/\psi,\psi(\mathrm{2S}),\Upsilon$, have been measured over the past few decades, for example see Refs.~\cite{H1:5,ZEUS:5,Aaltonen:2009kg,LHCb:1,LHCb:2,Aaij:2015kea}, with more data currently under analysis. The data measured at HERA sample the gluon distribution down to $x \sim 10^{-3}$ and data measured by the LHCb collaboration are sensitive to the gluon distribution down to $x \sim 10^{-5}$. However, the exclusive HVM data are not typically included in the global PDF fits. The main reason for this is that the theoretical frameworks used to describe exclusive HVM production differ to varying degrees from the standard framework used to describe more inclusive processes. 

In \sect{sec:HVM} we give a brief overview of the description of exclusive HVM production in both the $k_T$ and collinear factorisation frameworks. Our attempts to extract a gluon distribution from the measurements of this process at HERA and the LHC using $k_T$ factorisation are described in \sect{sec:ktfac}. The process is discussed using the collinear factorisation approach at NLO in \sect{sec:collinear}.

\section{Overview of Exclusive Heavy Vector Meson Production} \label{sec:HVM}

The various frameworks describing exclusive HVM production differ in two key ways from that typically used by global analyses. 

Firstly, the momentum of the incoming hadron need not be equal to the momentum of the outgoing hadron, therefore, the parton distributions probed are not the conventional PDFs. In the collinear factorisation framework exclusive processes such as GDVCS~\cite{Radyushkin:1996nd,Ji:1996nm,Collins:1998be} and exclusive vector meson electroproduction\footnote{This process is also known as deeply virtual meson production (DVMP).} (including HVM electroproduction)~\cite{Radyushkin:1996ru,Collins:1996fb} are described at amplitude level in terms of a convolution of coefficient functions and Generalized Parton Distributions (GPDs). These distributions and their relation to the PDFs are briefly described in \sect{sec:gpds}. Note that in this framework exclusive HVM photoproduction has not been proven to factorise, although, at least at NLO this appears to be the case~\cite{Ivan}. In the $k_T$ factorisation framework the process is typically described in terms of so-called `skewed unintegrated' PDFs~\cite{Martin:2001ms,Watt:2003vf}.

Secondly, the framework must describe the formation of the HVM. In this work, we use the leading term of the non-relativistic QCD (NRQCD) approach, it depends on an NRQCD matrix element $\langle O_1 \rangle$ which is proportional to the vector meson branching fraction to leptons $\Gamma \left[ V \rightarrow e^+ e^- \right]$~\cite{Bodwin:1994jh}. To this order the description is equivalent to the `static' approximation~\cite{Berger:1980ni}. 
The formation of the HVM is subject to modification by relativistic corrections. These corrections have been studied in a framework similar to NRQCD and found to be small~\cite{Hoodbhoy:1996zg}, they are neglected here. Note, however, that other work, based on integrating over the relative momentum of the outgoing heavy quarks, indicates that the relativistic corrections may be large~\cite{Frankfurt:1997fj}.

For ultraperipheral production there is an additional complication. In this process two protons (or heavy ions) collide. At LO, considering only hard perturbative contributions and neglecting background processes in which the protons break up or enter an excited state, this process can be modelled in terms of HVM photoproduction. The Equivalent Photon Approximation (EPA), utilised here, is widely used. It allows the ultraperipheral cross-section to be written as
\begin{equation}
\frac{\mathrm{d} \sigma(pp)}{\mathrm{d} y} = S_+^2 (y) N_+ \sigma_+(\gamma p) + S_-^2(y) N_- \sigma_-(\gamma p) + \ldots.
\end{equation}
Here $y$ is the rapidity, $\sigma_+(\gamma p)$, $\sigma_-(\gamma p)$ are photoproduction cross-sections computed with different $\gamma p$ centre-of-mass energies depending on the rapidity, $S_+^2(y)$, $S_-^2(y)$ are survival factors and $N_+$, $N_-$ are photon fluxes. Note that, without forward proton tagging, it is not possible to distinguish which proton emitted the photon entering the photoproduction amplitude. Therefore, the ultraperipheral cross-section for a given rapidity depends on the photoproduction cross-section evaluated at two $\gamma p$ centre-of-mass energies. The higher energy photoproduction cross-section accounts for photon emission from the forward proton whilst the lower energy photoproduction cross-section accounts for photon emission from the backward proton. The survival factors describe the probability that the rapidity gaps survive soft re-scattering effects between the interacting hadrons. They must be modelled and fitted to data, for this we use the Khoze-Martin-Ryskin model~\cite{Khoze:2002dc}. The photon flux used in this work is described in Ref.~\cite{Jones:2013pga}. The dependence on the photon flux $N_+$, $N_-$ partly cancels against the photon flux appearing in the denominator of the survival factors and so the same photon flux should be used to compute both quantities. The ellipses denote interference terms which are small and are neglected here.

Finally, especially for $J/\psi$ photoproduction, we have a low central scale (squared) of $2.4$~$\mathrm{GeV}^2$ and could reasonably question the validity of a perturbative approach. One might expect poor convergence of the pertubative series (due to large $\alpha_S$) and large effects not captured by perturbation theory. Indeed, NLO results in the collinear factorisation framework seem to indicate a very poor convergence of the perturbative series for reasonable choices of the renormalization and factorisation scales, see \sect{sec:collinear}.

\subsection{Generalized Parton Distributions} \label{sec:gpds}


Analogously to the PDFs, the GPDs can be defined through the matrix elements of quark and gluon operators at a light-like separation. For GPDs, unlike for PDFs, the momentum $p$ of the incoming hadron need not be equal to the momentum $p^\prime$ of the outgoing hadron. The GPDs $\mathcal{H}_g(x,\xi,t)$, $\mathcal{H}_q(x,\xi,t)$ depend on the momentum fractions $x$, $\xi$ of $P^+ = ({p}^+ + {p^\prime}^+) /2$ carried by the interacting partons. Conventionally, the incoming parton carries momentum $(x+\xi)P^+$ and the outgoing parton carries momentum $(x-\xi)P^+$, see the right panel of \fig{fig:setup}. The GPDs depend also on the Mandelstam variable $t=(p - p^\prime)^2$ and, after mass factorisation, on the factorisation scale $\mu_F^2$ which is suppressed here. In the forward limit, defined by $p=p^\prime$ or equivalently $\xi = 0$ and $t=0$, the GPDs are equal to the PDFs,
\begin{align}
\mathcal{H}_q(x,0,0) = & q(x), \quad x>0, \\
\mathcal{H}_q(x,0,0) = & -\bar{q}(-x), \quad x<0, \\
\mathcal{H}_g(x,0,0) = & xg(x).
\end{align}
Efforts by several groups to directly extract GPDs from GDVCS data are on-going~\cite{Kumericki:2016ehc,Boer:2014kya,GonzalezHernandez:2012jv}. There also exist GPDs $\mathcal{E}_q(x,\xi,t)$ and $\mathcal{E}_g(x,\xi,t)$ which appear multiplied by the momentum transfer $(p^\prime - p)$, here we neglect them. In \sect{sec:collinear} we will consider the quantities $F_i(x,\xi,t)$ at $t=0$ which, neglecting the proton mass, are related to the GPDs as $F_i(x,\xi,0) = \sqrt{1-\xi^2} \mathcal{H}_i(x,\xi,0)$ for $i=q,g$.

It has been conjectured that in the small $x$ and $\xi$ limit, beyond the strict forward limit, the GPDs are related to the PDFs via the Shuvaev transform~\cite{Shuvaev:1999fm}. This is based on the observation that the anomalous dimensions of the Gegenbauer moments $H_N$ of the GPDs are equal to the anomalous dimensions of the Mellin moments $M_N$ of the PDFs. The polynomiality property of the GPDs $ H_N = \sum_{k=0}^{\lfloor (N+1)/2 \rfloor} c_k^N \xi^{2k}$ allows the Gegenbauer moments to be determined from the PDFs up to corrections $\mathcal{O}(\xi)$ at NLO. The Shuvaev transform reconstructs the GPDs from these Gegenbauer moments. 

Note that the Shuvaev transform strictly relates the GPDs to some effective forward parton distribution functions and the conjecture consists of equating these distribution functions to the PDFs. The equivalence requires that there are no singularities in the right half $N$-plane of the PDF input distributions~\cite{Martin:2008tm}. Further, the transform does not relate the GPDs to the PDFs in the $|x|<\xi$ (time-like) region. In this work we assume that the Shuvaev transform can be used to relate the GPDs to the PDFs in the $\xi < |x|$ (space-like) region.

\section{The $k_T$ factorisation approach} \label{sec:ktfac}

\begin{figure}
\begin{center}
\includegraphics[width=.4\textwidth]{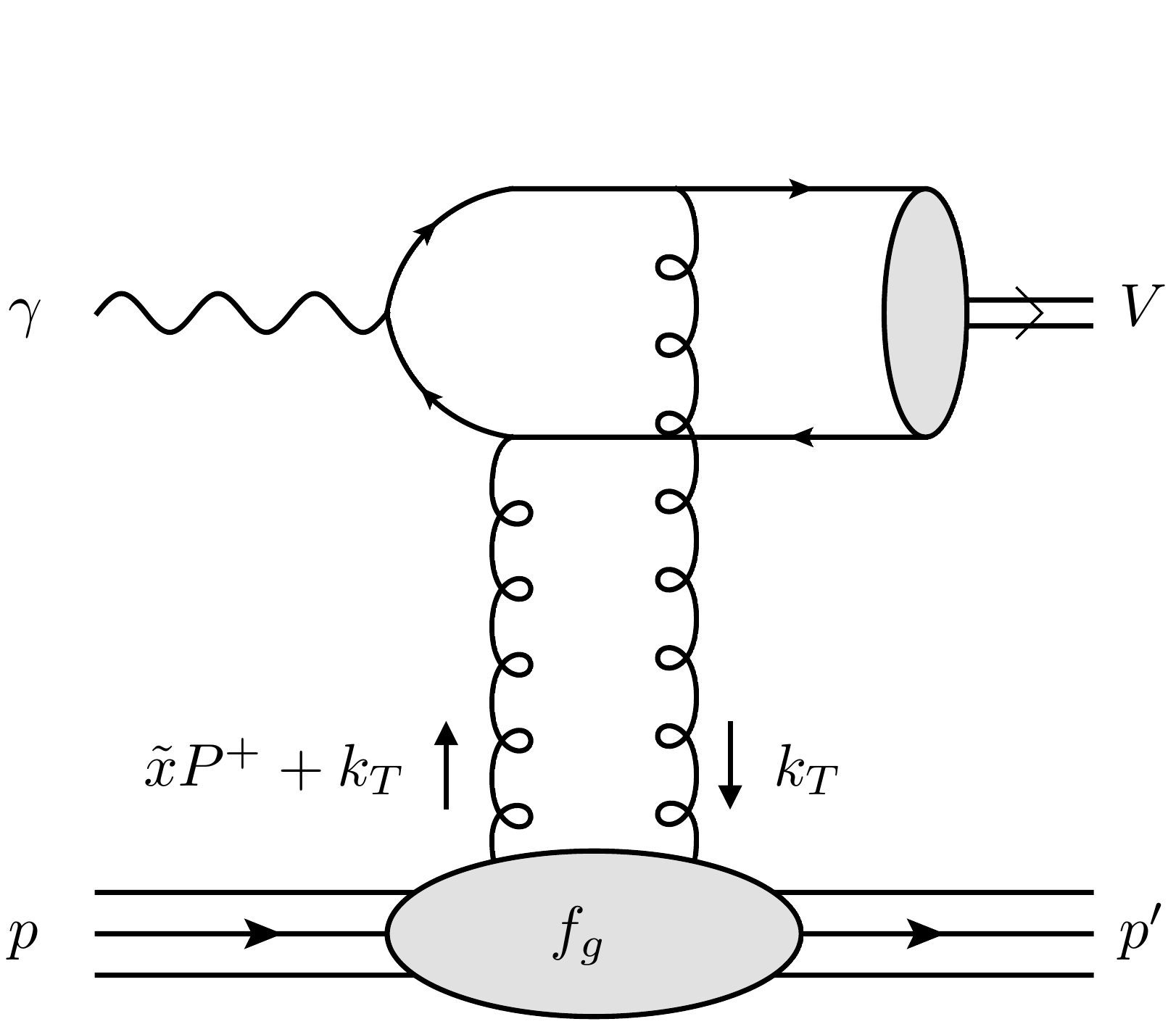}
\includegraphics[width=.4\textwidth]{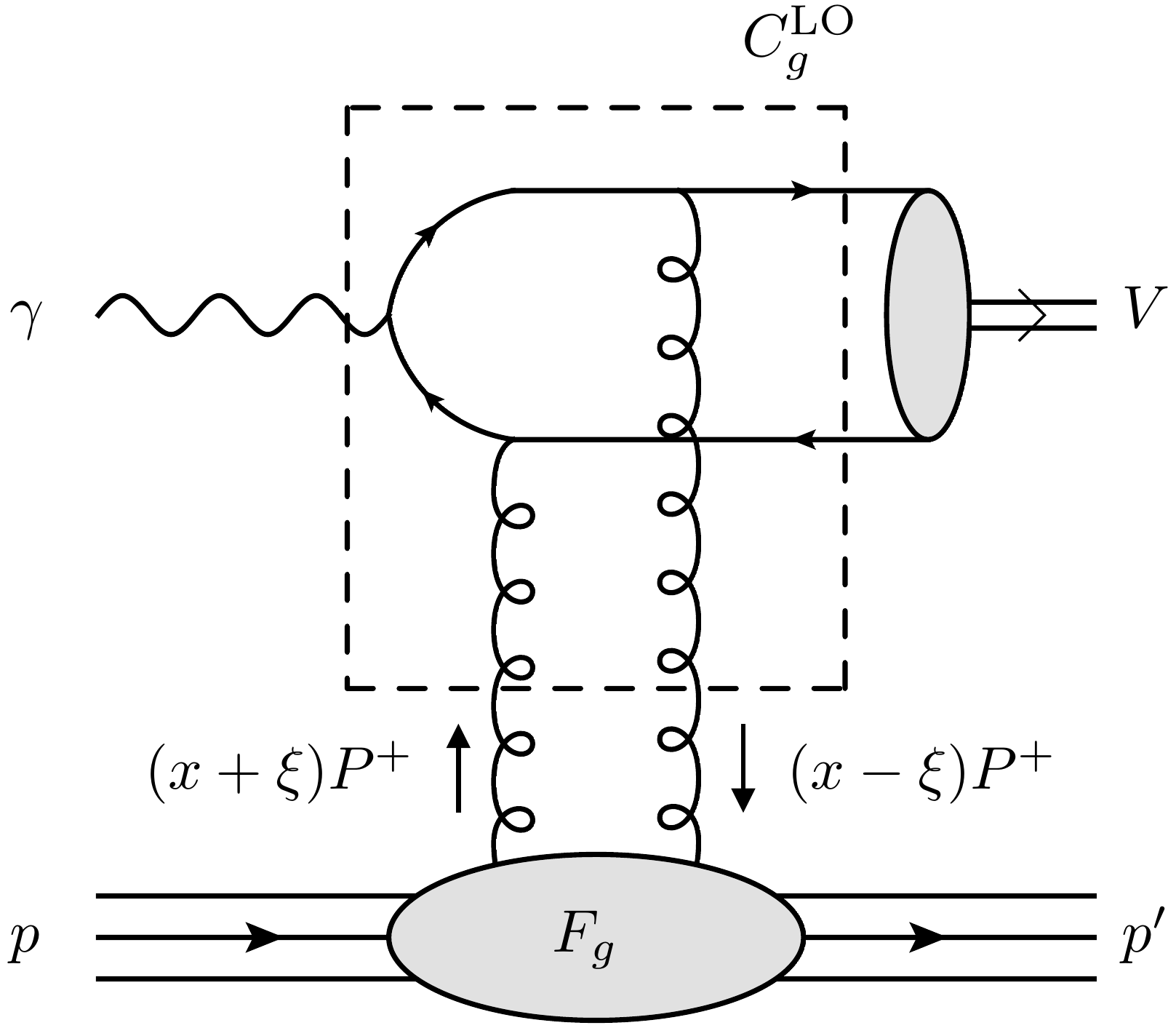}
\end{center}
\caption{Framework used in (left panel) our $k_T$ factorisation approach and (right panel) the collinear factorisation approach. The quarks connected to the photon are the massive ($c$ or $b$) valance quarks which form the vector meson $V=J/\psi,\psi(\mathrm{2S}),\Upsilon$. The dashed box in the right panel encloses a (5-leg) diagram that contributes to the coefficient function $C_g^{LO}$.}
\label{fig:setup}
\end{figure}

Previously we have studied exclusive HVM production in the context of $k_T$ factorisation~\cite{Jones:2013pga,Jones:2013eda}. In our work we calculated at tree level (LO) the dominant imaginary part of the amplitude for photoproduction and electroproduction. At tree level only gluon exchange between the proton and photon contributes. We considered only the high energy `maximal skew' approximation in which the incoming gluon carries plus momentum fraction $\tilde{x} \approx 2 x = 2 \xi$ and the outgoing gluon carries no momentum in the plus direction. In $k_T$ factorisation the incoming and outgoing gluons can be off-shell and have momentum $k_T$ transverse to the plus direction, see the left panel of \fig{fig:setup}.

In this framework the `skewed unintegrated' gluon distribution was parametrised in terms of a gluon PDF and a `skewing factor', which can be derived from the Shuvaev transform using the `maximal skew' approximation and assuming the gluon PDF obeys a simple power law form $xg(x,\mu_F^2) \sim x^{\lambda(\mu_F^2)}$. Note that in Ref.~\cite{Harland-Lang:2013xba} it was shown that the parametrisation used in our work can introduce an $\mathcal{O}(20-30\%)$ effect on the total cross-section as compared to using the full Shuvaev transform. Under the assumption that the Shuvaev transform is valid, that non-relativistic corrections to the HVM formation are small and neglecting contributions higher order in $\alpha_S$, this is the dominant uncertainty in our analysis.

Two models were considered. In the first model the approximation $k_T^2 \ll \bar{Q}^2$ was used along with a simple power law form for the gluon PDF. In the second model the full $k_T$ dependence was maintained and a double-leading logarithm approximation was used for the gluon PDF which mimics DGLAP evolution over the relevant small range of scales. Maintaining the full $k_T$ dependence in the second model includes some contributions that would only appear beyond LO in collinear factorisation. In both models the gluon PDF was parametrised with 3 variables that were fitted to HERA $J/\psi$ data~\cite{H1:5,ZEUS:5} (both photoproduction and electroproduction) and ultraperipheral data from CDF~\cite{Aaltonen:2009kg} and the LHCb 2013 data~\cite{LHCb:1}, but no other data were included in these fits. The fits obtained are in good agreement with the data, for the first model we obtain $\chi^2/\mathrm{d.o.f} = 41/79$ and for the second model we obtain $\chi^2/\mathrm{d.o.f} = 50/79$.
The extracted gluons are broadly similar to the global PDFs but the uncertainty bands at low $x \sim 10^{-5}$ for scales 2-- 6 $\mathrm{GeV}^2$ are much narrower.


Further, the predictions from these models compare well to subsequent data on $J/\psi$, $\psi(\mathrm{2S})$ and $\Upsilon$ measured at LHCb after our publications~\cite{LHCb:2,Aaij:2015kea}. Note that our $\Upsilon$ prediction does not rely on any $\Upsilon$ data but is just the result obtained from (approximate) DGLAP evolution of the gluon fitted from $J/\psi$ photoproduction and electroproduction. 

Unfortunately, going beyond tree level in this framework appears not to be straightforward, further, the connection between the `skewed unintegrated' distribution is less direct than in the collinear factorisation framework.

\section{The Collinear Factorisation Approach at NLO} \label{sec:collinear}

More recently we considered exclusive HVM photoproduction at NLO in collinear factorisation. In this framework the NLO result has already been known for more than a decade~\cite{Ivan} but contained errors corrected only recently~\cite{NocklesThesis2009,Ivan}. Our re-calculation, which confirms the corrected result, uses a different technique than the original authors, it is described in \sect{sec:calculation}.

At NLO exclusive HVM production receives additional contributions from a quark exchange between the proton and photon. In the collinear factorisation framework the amplitude can be written as
\begin{equation}
A \propto \int_{-1}^1 \mathrm{d} x \left[ C_g(x,\xi) F_g(x,\xi) + \sum_{q=u,d,s} C_q(x,\xi) F_q(x,\xi) \right], \label{eq:amplitude}
\end{equation}
where $C_g$, $C_q$ are perturbatively calculable coefficient functions and $F_g$, $F_q$ depend on the GPDs. Here the dependence of the coefficient functions on the renormalization scale $\mu_R^2$ and the factorisation scale $\mu_F^2$ is suppressed. The dependence of $F_g$, $F_q$ on the Mandelstam $t$ and $\mu_F^2$ has also been suppressed.

\subsection{Calculation} \label{sec:calculation}

To obtain the LO coefficient function we must compute the 5-leg diagrams contained in the dashed box of \fig{fig:setup}. The Lorentz or spin indices of the light partons connected to the GPD are contracted with GPD projectors whilst the heavy quarks which enter the NRQCD matrix element are contracted with a projector onto the HVM state, see Refs.~\cite{Ivan,Braaten:2002fi}. There are 6 gluon diagrams at tree level and 97 gluon + 12 quark diagrams at 1-loop. We compute using dimensional regularisation, the diagrams are generated using \texttt{QGRAF}~\cite{Nogueira:1991ex} then processed using \texttt{FORM}~\cite{Kuipers:2012rf}. The factorisation imposes that the momenta of the incoming and outgoing light partons are proportional whilst the leading term of the NRQCD expansion imposes that the momenta of the outgoing heavy quarks are equal. The na\"ive application of standard loop techniques will therefore lead to a singular decomposition and/or an incomplete reduction to the master integrals. 

To circumvent these problems we perform a Sudakov decomposition of all external momenta $p_i^\mu = (p_i \cdot n) p^\mu + (p_i \cdot p) n^\mu + {p_i}_T^\mu$. Due to the kinematic constraints we see that all transverse components ${p_i}_T$ can be chosen to vanish. All external momenta can therefore be constructed from just two linearly independent vectors which can be chosen as the Sudakov basis vectors $p$, $n$. Thus, we can decompose our integrals in terms of just $p$, $n$ (and the metric). 

Further, the linear dependence of the external momenta implies that there are relations between the propagators. These relations allow us to reduce $N$ point integrals to $N-1$ point integrals by eliminating linearly dependent propagators from each integral. 
Since the kinematics of the process impose that at most two external momenta can be linearly independent there can be at most three different propagators appearing in any integral. 

After eliminating the linear dependencies and using integration by parts identities, as implemented in \texttt{REDUZE}~\cite{Studerus:2009ye}, we obtain scalar triangle, bubble and tadpole master integrals only. These integrals depend on only one scaleless ratio. The analytic continuation of our result for the coefficient functions is checked against the numerical implementation of the integrals in Ref.~\cite{Hahn:1998yk}.

\subsection{Results}

\begin{figure}
\begin{center}
\includegraphics[width=.35\textwidth,angle=-90]{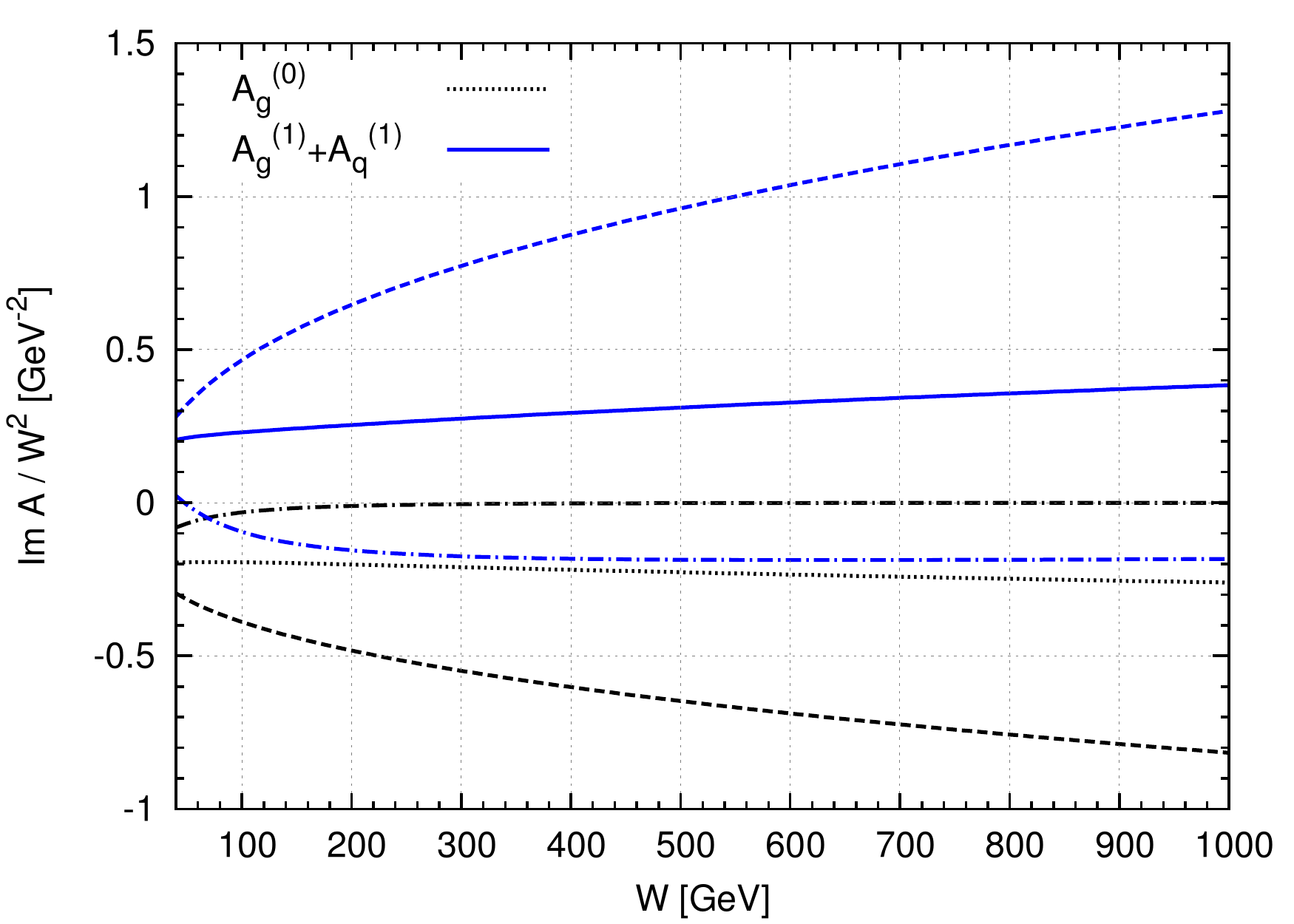}
\includegraphics[width=.35\textwidth,angle=-90]{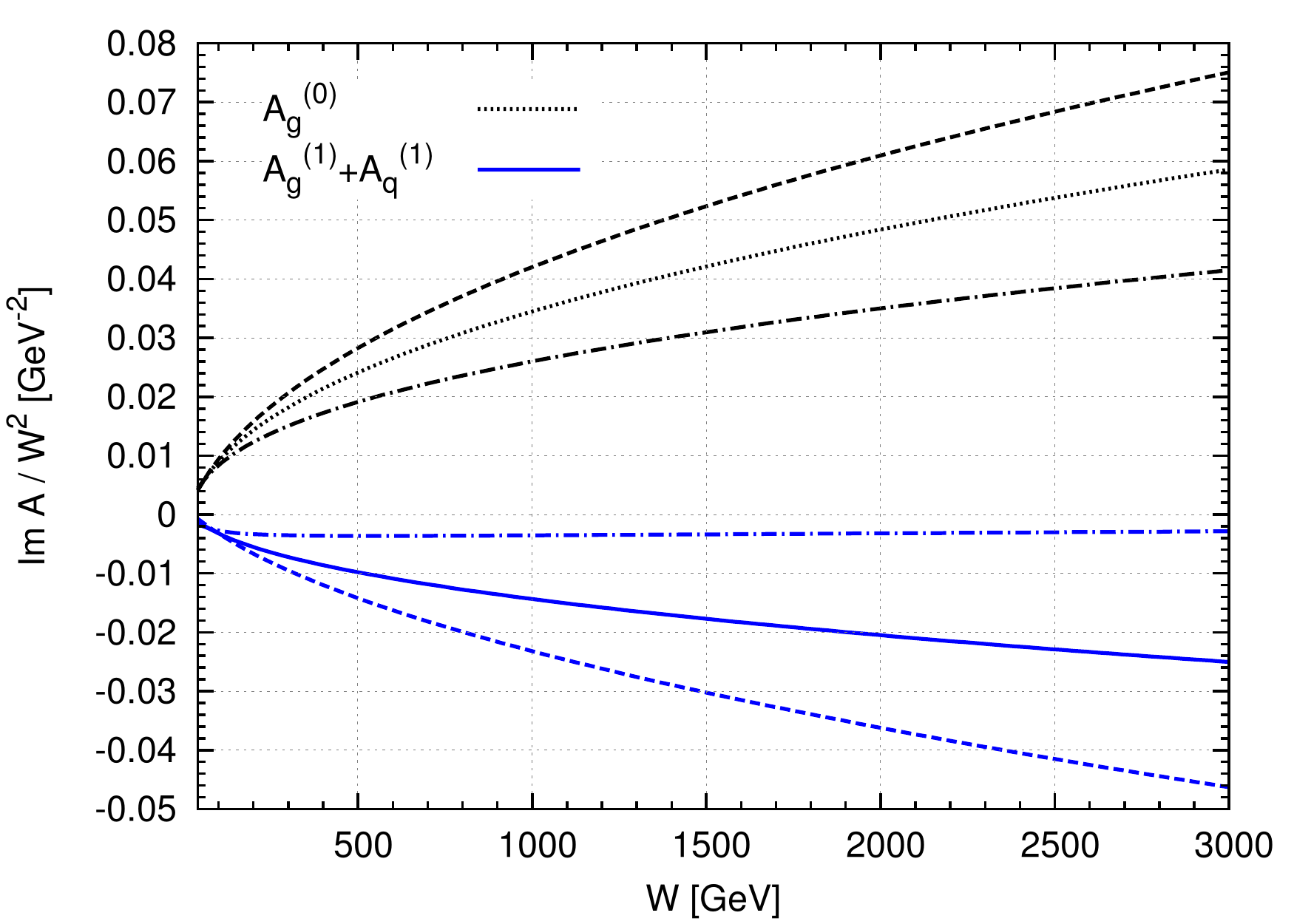}
\end{center}
\caption{The imaginary part of the born and 1-loop contributions to the amplitude as a function of the $\gamma p$ centre-of-mass energy $W$ for (left panel) $\gamma p \rightarrow J/\psi + p$ and (right panel) $\gamma p \rightarrow \Upsilon + p$ produced using CTEQ6.6 partons~\cite{Nadolsky:2008zw}. The bands show the $\mu^2 \equiv \mu_R^2 = \mu_F^2$ variation about the central scale. For the left panel $\mu^2 = 1.2,2.4,4.8$~$\mathrm{GeV}^2$ and for the right panel $\mu^2 = 11.9,22.4,44.7$~$\mathrm{GeV}^2$. The dot-dashed, solid and dashed lines correspond to the low, central and high values of the scale $\mu$, respectively.}
\label{fig:me}
\end{figure}

To obtain the amplitude the coefficient functions calculated as described in \sect{sec:calculation} must be convoluted with the GPDs according to Eq.~\eqref{eq:amplitude}. Rather than fitting a gluon to the $J/\psi$ data as in \sect{sec:ktfac} we instead extract GPDs from the global PDFs using the full Shuvaev transform implemented as described in Ref.~\cite{Martin:2008tm}. Since the Shuvaev transform limits us to the $\xi < |x|$ space-like region we can consider only the imaginary part of the amplitude which is zero outside this region (this can be seen explicitly from the coefficient functions). Further, it should be noted that, even if valid, the Shuvaev transform may be a poor approximation for large $\xi$ (small $\gamma p$ centre-of-mass energy $W$). The result for the born and sum of quark and gluon 1-loop contributions to the amplitude for $J/\psi$ and $\Upsilon$ are shown in \fig{fig:me}. For the central value of the scale for the $J/\psi$ the 1-loop contributions dominate the born contribution and have the opposite sign which leads to a (nonsensical) negative cross-section. For $\Upsilon$ the situation is not as severe, for the central value of the scale the born contribution dominates the 1-loop correction leading to a positive cross-section which is, however, very sensitive to the scale choice (for $W$ greater than a few hundred $\mathrm{GeV}$ the scale variation band encompasses negative cross-sections). This poor perturbative stability is a major obstacle for any attempt to extract either GPDs or PDFs from exclusive HVM photoproduction.

Nevertheless, the sensitivity to the scale for $W \gg M_V^2$, where $M_V^2$ is the HVM mass, can be understood. It originates from terms containing $\ln(m^2/\mu_F^2) \ln(1/\xi)$, where $m$ is the heavy quark mass, generated by mass factorisation, which can become large for small $\xi$. It was argued in Ref.~\cite{Jones:2015nna} that a scale fixing procedure can resum these double log terms and reduce the sensitivity to the factorisation scale. Another approach being pursued is to resum the leading $\ln(1/\xi)$ \`a la BFKL~\cite{Ivanov:2015hca}. Furthermore, for $J/\psi$ one might reasonably suspect that the low central scale is at least partly responsible for the poor perturbative stability.

\section{Conclusions}

We have described our LO results for exclusive HVM in the $k_T$ factorisation framework. In this framework the gluon distribution extracted from $J/\psi$ exclusive and ultraperipheral production has a reduced uncertainty at low $x\sim 10^{-5}$ and low scales 2 -- 6 $\mathrm{GeV}^2$ as compared to the global analyses~\cite{Jones:2013pga,Jones:2013eda}.

Recently we have also re-calculated and confirmed the (recently corrected) NLO result for exclusive photoproduction~\cite{Ivan}. For both the $J/\psi$ and the $\Upsilon$, the 1-loop contribution is large, comparable to the born contribution, and very sensitive to the factorisation scale choice. The loop contribution has opposite sign to the born contribution which can lead even to negative cross-sections for seemingly reasonable scale choices. For the $\Upsilon$, the correction is large but the cross-section is positive for $\gamma p$ centre-of-mass energies less than a few hundred $\mathrm{GeV}$. The poor behaviour of the correction at large centre-of-mass energies for both $J/\psi$ and $\Upsilon$ is at least partly due to the appearance of large $\ln(1/\xi)$ and several groups are working, with some success, to address this primarily by attempting to resum these logarithms~\cite{Jones:2015nna,Ivanov:2015hca}.

\section*{Acknowledgements}
SPJ is supported by the Research Executive Agency (REA) of the European Union under the Grant Agreement PITN-GA2012316704 (HiggsTools), MGR is supported by the RSCF grant 14-02-00281 and TT is supported by STFC under the consolidated grant ST/L000431/1.

\end{document}